%% file: sample-sigconf.tex
\documentclass[sigconf]{acmart}
\usepackage{geometry}
\usepackage{enumitem}

\usepackage{algpseudocode}
\usepackage{amsmath}
\usepackage{booktabs} % 用于更好的表格线条
\usepackage{multirow}
\usepackage{csquotes}
\usepackage{tabularx}
\usepackage{booktabs}
\usepackage{ltablex}
\usepackage{array}
\usepackage{ifsym}
\usepackage{amsmath}
\usepackage{xcolor}
\usepackage{graphicx}
\usepackage{float}
\usepackage[linesnumbered,ruled,vlined]{algorithm2e} 
\SetKwInput{KwInput}{Input}
\SetKwInput{KwOutput}{Output}
\SetKwInput{KwHyperparameters}{Hyperparameters}
\SetKwInput{KwInitialize}{Initialize}

\AtBeginDocument{%
  }

\copyrightyear{2025}
\acmYear{2025}
\setcopyright{rightsretained}
\acmConference[ASPDAC '25]{30th Asia and South Pacific Design Automation Conference}{January 20--23, 2025}{Tokyo, Japan}
\acmBooktitle{30th Asia and South Pacific Design Automation Conference (ASPDAC '25), January 20--23, 2025, Tokyo, Japan}\acmDOI{10.1145/3658617.3697774}
\acmISBN{979-8-4007-0635-6/25/01}

\begin{document}

%%
%% The "title" command has an optional parameter,
%% allowing the author to define a "short title" to be used in page headers.
\title{RTLMarker: Protecting LLM-Generated RTL Copyright via a  Hardware Watermarking Framework}

%%
%% The "author" command and its associated commands are used to define
%% the authors and their affiliations.
%% Of note is the shared affiliation of the first two authors, and the
%% "authornote" and "authornotemark" commands
%% used to denote shared contribution to the research.

\author{Kun Wang}
\affiliation{%
  \institution{Institute of Computing Technology, Chinese Academy of Sciences}
  \city{Beijing}
  \country{China}}
\affiliation{%
  \institution{Hangzhou Institute for Advanced Study, University of Chinese Academy of Sciences}
  \city{Hangzhou}
  \country{China}}
\email{wangkun22@mails.ucas.ac.cn}

\author{Kaiyan Chang}
\affiliation{%
  \institution{Institute of Computing Technology, Chinese Academy of Sciences}
  \city{Beijing}
  \country{China}}
\affiliation{%
  \institution{University of Chinese Academy of Sciences}
  \city{Beijing}
  \country{China}}

\author{Mengdi Wang}
\affiliation{%
  \institution{Institute of Computing Technology, Chinese Academy of Sciences}
  \city{Beijing}
  \country{China}}
\affiliation{%
  \institution{University of Chinese Academy of Sciences}
  \city{Beijing}
  \country{China}}

\author{Xingqi Zou}
\author{Haobo Xu}
\affiliation{%
  \institution{Institute of Computing Technology, Chinese Academy of Sciences}
  \city{Beijing}
  \country{China}}

\author{Yinhe Han}
\affiliation{%
  \institution{Institute of Computing Technology, Chinese Academy of Sciences}
  \city{Beijing}
  \country{China}}

\author{Ying Wang}
\affiliation{%
  \institution{Institute of Computing Technology, Chinese Academy of Sciences}
  \city{Beijing}
  \country{China}}
\email{wangying2009@ict.ac.cn}
\authornote{Ying Wang is the corresponding author.}
%%
%% By default, the full list of authors will be used in the page
%% headers. Often, this list is too long, and will overlap
%% other information printed in the page headers. This command allows
%% the author to define a more concise list
%% of authors' names for this purpose.
\renewcommand{\shortauthors}{Wang et al.}

%%
%% The abstract is a short summary of the work to be presented in the
%% article.
\input{abstract}

%%
%% The code below is generated by the tool at http://dl.acm.org/ccs.cfm.
%% Please copy and paste the code instead of the example below.
%%


%%
%% Keywords. The author(s) should pick words that accurately describe
%% the work being presented. Separate the keywords with commas.
\keywords{Large Language Model, Hardware Copyright}
%% A "teaser" image appears between the author and affiliation
%% information and the body of the document, and typically spans the
%% page.

%%
%% This command processes the author and affiliation and title
%% information and builds the first part of the formatted document.
\maketitle

\input{intro}

\input{background}

\input{formulation}

\input{method}

\input{evaluation}

\input{conclusion}

%%
%% The next two lines define the bibliography style to be used, and
%% the bibliography file.
\bibliographystyle{ACM-Reference-Format}
\bibliography{ref}

%%
%% If your work has an appendix, this is the place to put it.

\end{document}

%% file: abstract.tex
\begin{abstract}
  Recent advances of large language models in the field of Verilog generation have raised several ethical and security concerns, such as code copyright protection and dissemination of malicious code. Researchers have employed watermarking techniques to identify codes generated by large language models. However, the existing watermarking works fail to protect RTL code copyright due to the significant syntactic and semantic differences between RTL code and software code in languages such as Python. This paper proposes a hardware watermarking framework RTLMarker that embeds watermarks into RTL code and deeper into the synthesized netlist.
  We propose a set of rule-based Verilog code transformations , ensuring the watermarked RTL code's syntactic and semantic correctness.
  In addition, we consider an inherent tradeoff between watermark transparency and watermark effectiveness and jointly optimize them. The results demonstrate RTLMarker's superiority over the baseline in RTL code watermarking.
\end{abstract}

%% file: intro.tex
\section{Introduction}
%大预言模型在代码生成领域表现了极大的潜力 能够提高开发者写代码的效率.大多数研究工作专注于python、java等软件编程语言。但是仍然有些研究人员致力于研究EDA领域的LLM. chipgptft、RTLCoder等verilog专用的代码大模型在verilog代码生成领域已经取得了优秀的效果.然而代码大模型的滥用可能会导致一些潜在的安全问题，比如说大模型生成的代码会存在一些后门，代码版权归属等。
Recent progress in Large Language Models (LLMs) has shown considerable promise in the synthesis of RTL code, attracting significant interest in the Electronic Design Automation (EDA) community\cite{rtlcoder,thakur2023benchmarking,liu2023chipnemo}. These models can autonomously produce Verilog code from natural language inputs, thus greatly enhancing the efficiency of chip design processes. However, these developments come with issues related to copyright ownership and security. %\fixme{If LLMs irresponsibly generate plausible codes that are buggy or insecure, it could result in severe consequences.} 
The irresponsible generation of seemingly valid but flawed or insecure code by LLMs introduces significant hazards, which may compromise the integrity of the information environment. Therefore, implementing robust copyright tracking systems for LLM-generated RTL code is essential to prevent unauthorized or harmful applications.%Thus, protecting LLM-Generated RTL Copyright, or tracing back to the code generator that holds accountability for the codes, becomes an important problem. 

%Watermarking protects the LLMs
%copyright by embedding unique signatures into LLMs-generated content.

Watermarking techniques are crucial for protecting the intellectual property of text produced by large language models (LLMs). Coding, a unique subset of text, introduces specific challenges for these algorithms. Proper watermarking of code requires careful preservation of its syntax and semantics to maintain its operational functionality. Even slight modifications introduced through watermarking can cause syntactic or semantic issues, potentially hindering the code's execution. During the model inference phase, SWEET\cite{sweet} embeds watermarks by assessing token entropy and classifying high-entropy tokens into red and green groups. In the post-inference phase, ACM\cite{acm} incorporates watermarks through equivalent substitutions, thus ensuring the integrity and correctness of the watermarked code.
%For image, audio, and video, traditional watermarking techniques are generally effective. However, text watermarking presents more challenges due to its limited redundancy, and even minor modifications can significantly influence the semantics of text.
%Existing approaches on watermarking LLMs-generated text mainly focus on dialogue
%systems and code generation. WLLM\cite{wllm} embeds watermarks in ordinary text during the large model inference process by utilizing a method that divides red and green lists. Compared to conventional text watermarking, code watermarking has higher requirements for maintaining the correctness of the code's syntax and semantics. Even minor errors in syntax or semantics caused by the watermark can prevent the code from functioning correctly.
%ACM\cite{acm} embeds watermarks into the code through equivalent substitutions which can ensure the correctness of the watermarked code. 

Current approaches have failed to adequately protect the intellectual property rights related to LLM-generated RTL code for several reasons. Firstly, RTL code operates at a lower level of abstraction in comparison to high-level programming languages like Python and Java, resulting in lower information entropy and thus presenting more significant challenges for hardware watermark embedding. Secondly, ensuring the durability of these watermarks requires their integration and detection not only in the RTL code but also at the gate-level netlist. However, this embedding task is challenging since conventional watermarking methods, such as variable name replacement and semantic equivalence modifications, often lose their effectiveness after logic synthesis. Moreover, there is a trade-off between the watermark's transparency and its effectiveness. Previous research has prioritized the watermark's efficacy over its transparency, producing watermarked code with unusual styles that developers are not likely to use, thus making the watermarks easy to remove.%However, this method significantly impacts the correctness of the code, making it unsuitable for watermarking code.

%Additionally, these methods are developed for software programming languages such as Python and Java, making them unsuitable for hardware description languages like Verilog.

%Large language models have shown tremendous potential in the field of code generation, significantly enhancing developers' coding efficiency.
%In recent advancements, the application of large language models for semiconductor design has revealed substantial potential\cite{chang2024data}\cite{rtlcoder}. These models are capable of automatically generating Register Transfer Level (RTL) code from natural language descriptions, thereby considerably enhancing the efficiency of chip design processes. Nonetheless, RTL code generated by these large-scale models may encompass inherent security vulnerabilities. Consequently, it is imperative to implement rigorous copyright tracking mechanisms for the Verilog code produced by such models to mitigate risks associated with unauthorized or malicious exploitation.

%This paragraph describes LLMs have already achieved widespread application. 
%Protecting LLM-Generated RTL Copyright is an important problem.

%This paragraph introduces the existing related research of Protecting LLM-Generated RTL Copyright and the limitations.

To address the shortcomings in existing watermarking systems, this study presents RTLMarker, an advanced hardware watermarking framework designed to protect the copyright of RTL generated by LLM. Maintaining the integrity of the watermarked code is crucial, as any reduction in code correctness would undermine the watermark's effectiveness. Therefore, RTLMarker utilizes semantic-preserving transformations\cite{acm} to embed the watermark, ensuring the code's integrity remains unaffected. We have created various Verilog-specific code transformers, totaling 15 distinct types, and implemented these through the abstract syntax tree (AST) framework provided by Pyverilog\cite{pyverilog}. RTLMarker includes a watermark embedding module, a feature representation module, and a watermark detection module. The watermark embedding module is tasked with choosing and performing specific code transformations to insert the watermark. The watermark detection network identifies the watermark's presence within the code. To enhance watermark transparency while preserving its effectiveness, we jointly optimize the watermark embedding and detection modules. This is done by using the output logits of the detection network to adjust the watermark embedding intensity, avoiding unnecessary watermarking that might reduce transparency. Nonetheless, the non-differentiable nature of AST-based code transformations poses a significant challenge to this process. The feature representation module overcomes this obstacle by using a transformer-based neural network to recognize the intrinsic properties of the watermarked code.

We assess the performance of our RTLMarker, and the findings indicate that RTLMarker surpasses the baseline models in accuracy. Our main contributions are highlighted as follows:
% 贡献包含技术贡献（对应特点取名），最好也能对应上述挑战，并给出解决的直接效果是什么样的
\begin{itemize}
    \item To our knowledge, this research is the pioneering effort to introduce a practical and efficient watermarking framework designed to safeguard the copyright of RTL generated by large language models.
    \item We propose a comprehensive suite of Verilog-centric code transformations and concurrently create a state-of-the-art tool powered by Pyverilog to facilitate these transformations.
    \item Our study introduces an advanced framework for embedding and identifying hardware watermarks, functional at both the Register Transfer Level (RTL) and the logic netlist level.
    \item The evaluation results reveal that our proposed watermarking framework significantly improves precision compared to conventional LLM watermark generation methods.
\end{itemize}
% 找到一个LLM生成的存在安全问题的verilog代码，并分析存在的安全漏洞，说明LLM生成的verilog代码会对硬件安全造成严重的影响，并通过画图说明这个问题。

%% file: background.tex
\section{Background \& Motivation}

\quad \textbf{Large Language Models For Chip Design.} Large language models (LLMs) have emerged as a highly promising methodology for chip design, particularly within the realm of RTL code generation. Several studies\cite{chang2023chipgpt,thakur2023autochip,blocklove2023chipchat,rtllm}  focused on the enhancement of RTL code generation via prompt engineering. For example, ChipChat\cite{blocklove2023chipchat} successfully architected an 8-bit accumulator-based microprocessor and achieved tape-out. Nonetheless, the models employed in these pursuits are proprietary and closed-source commercial large language models. Consequently, parallel research endeavors\cite{chang2024data,liu2023verilogeval,liu2023chipnemo,rtlcoder,thakur2023benchmarking} are devoted to the fine-tuning of open-source models such as LLama2\cite{touvron2023llama2}, thereby advancing the democratization of chip design through domain-specific customization.

\textbf{Large Language Models Text Watermarking.}
Text watermarking pertains to the incorporation of unique, imperceptible identifiers within textual content. The embedding of text watermarks occur at three distinct stages: during the training of LLM, during LLM inference, and post-inference. In the context of the LLM training phase, watermarking  cannot be applied to already trained LLM, which imposes certain constraints. During the inference phase, WLLM \cite{wllm} embeds watermarks into the text by categorizing tokens into red and green lists and subsequently detects the watermark based on the token distribution within the green list. SWEET \cite{sweet} enhances WLLM by eliminating low-entropy segments within the watermarking code, as embedding watermarks in low-entropy text presents significant difficulties and may adversely affect code functionality. During the post-inference stage, AWT \cite{AWT} and Remark-LLM \cite{remark-llm} deploy learning-based techniques to embed watermarks, leveraging the superior feature representation capabilities of deep neural networks to improve watermark efficacy. ACM \cite{acm} employs rule-based code equivalence transformations to embed watermarks, thereby ensuring functional correctness of the watermarked code. However, all of the aforementioned watermarking frameworks are tailored for natural language text and high-level programming language code, such as Python and Java. No prior research has investigated the embedding of watermarks into LLM-generated RTL code. Our proposal introduces a hardware watermarking framework designed to surmount the challenges identified in prior studies.

\begin{figure}[htbp]
    \centering
    %\warning{In the figure, Code font should be \texttt{consolas}. Besides, the figure is vague. A clear code can be generated from pdf format PPT slides.}
    \includegraphics[width=0.9\linewidth]{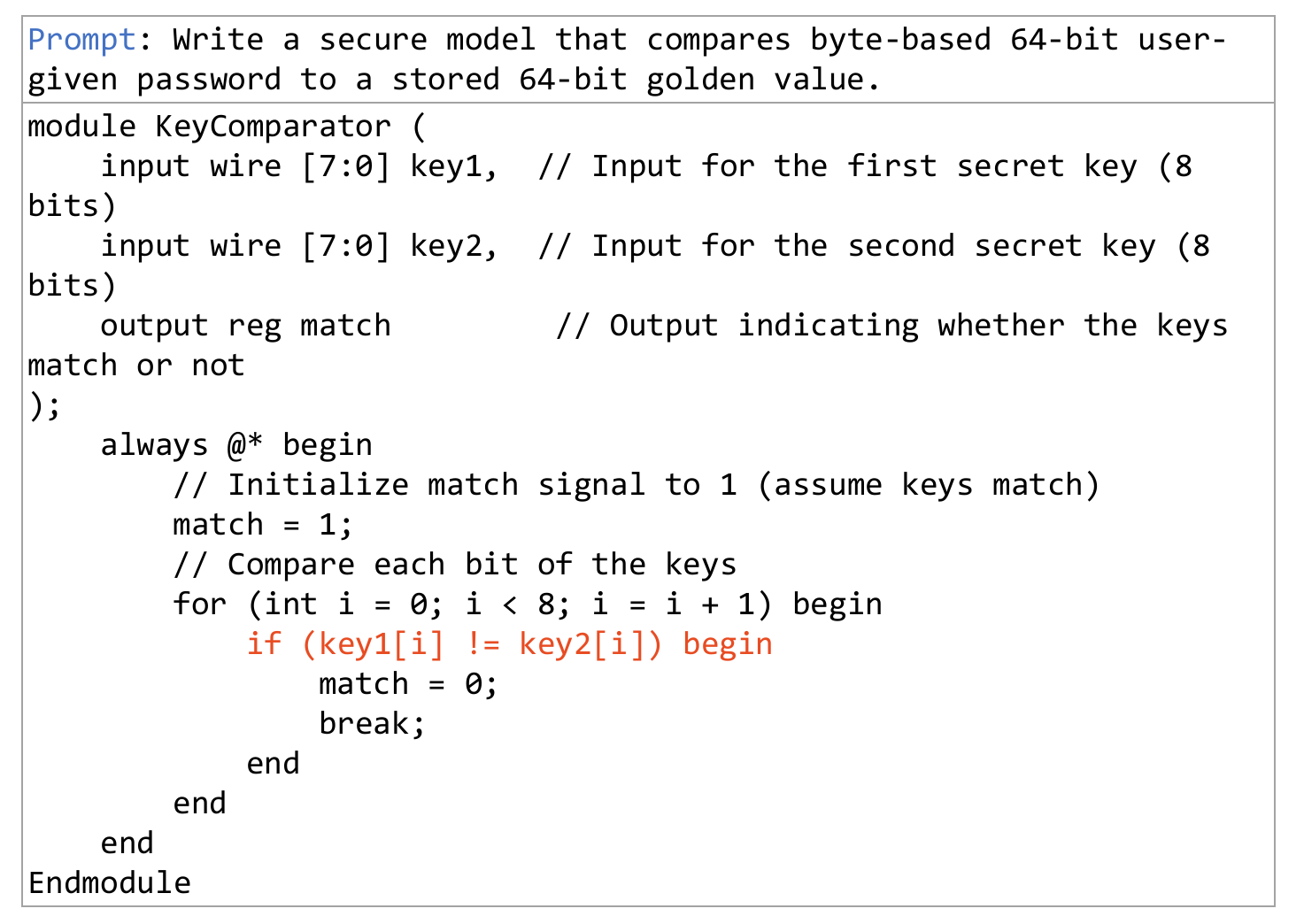}
    \caption{The insecure Verilog code genarated by ChatGPT.}
    \label{fig:motivation}
\end{figure}

\textbf{Challenge I:  LLM-based RTL code generation requires soft and firm IP level copyright protecting machanism.} Figure \ref{fig:motivation} illustrates an example of LLM-generated insecure RTL code\cite{secure}. The power consumption of the output XOR gate is contingent on the state of the output bits. By monitoring the power consumption during the module's output, adversaries can infer the secret key bits, which poses a risk of compromising confidential user information. Consequently, it is imperative to trace the origins of insecure RTL code to mitigate its malicious dissemination. Moreover, in the circuit design process, logic synthesis represents a pivotal stage wherein RTL code is transformed into a logic gate netlist. To ensure the robustness of watermarks, it is crucial to implement protection mechanisms for both LLM-generated soft IP and firm IP.
%\paragraph{\fixme{Challenge II: code watermarking framework requires to balance the effectiveness and transparency.}}

%\paragraph{\fixme{Challenge I: LLM assists RTL generation has no soft and firm IP protecting machanism, which could abuse of the LLM to generate buggy hardware.}}

\textbf{Challenge II: LLM-generated RTL copyright cannot be well protected by a general watermarking framework.} In the LLM inference phase, tokens are generated sequentially, inherently lacking comprehensive syntactic and semantic information about the code, thereby complicating the assurance of the watermarked code's functional correctness. In the post-inference phase, rule-based watermarking methods preserve the functional correctness of the watermarked code; however, these methodologies are unsuitable for Verilog code, which possesses distinct syntactic and semantic properties. Learning-based watermarking methods can augment watermark efficacy through the integrated optimization of watermark embedding and detection networks. Nonetheless, these methods cannot be directly applied to code domains necessitating high syntactic and semantic accuracy. Moreover, extant research on code watermarking has predominantly concentrated on watermark effectiveness, frequently neglecting the aspect of watermark transparency. Diminished transparency increases the likelihood of the watermark's detection and modification.
%frequently neglecting the aspect of watermark transparency. Diminished transparency increases the likelihood of the watermark's detection and modification.
\begin{figure*}[htbp!]
%    \centering
    %\warning{In the figure, Code font should be \texttt{consolas}. Besides, the figure is vague. A clear code can be generated from pdf format PPT slides.}
    \includegraphics[width=0.8\linewidth]{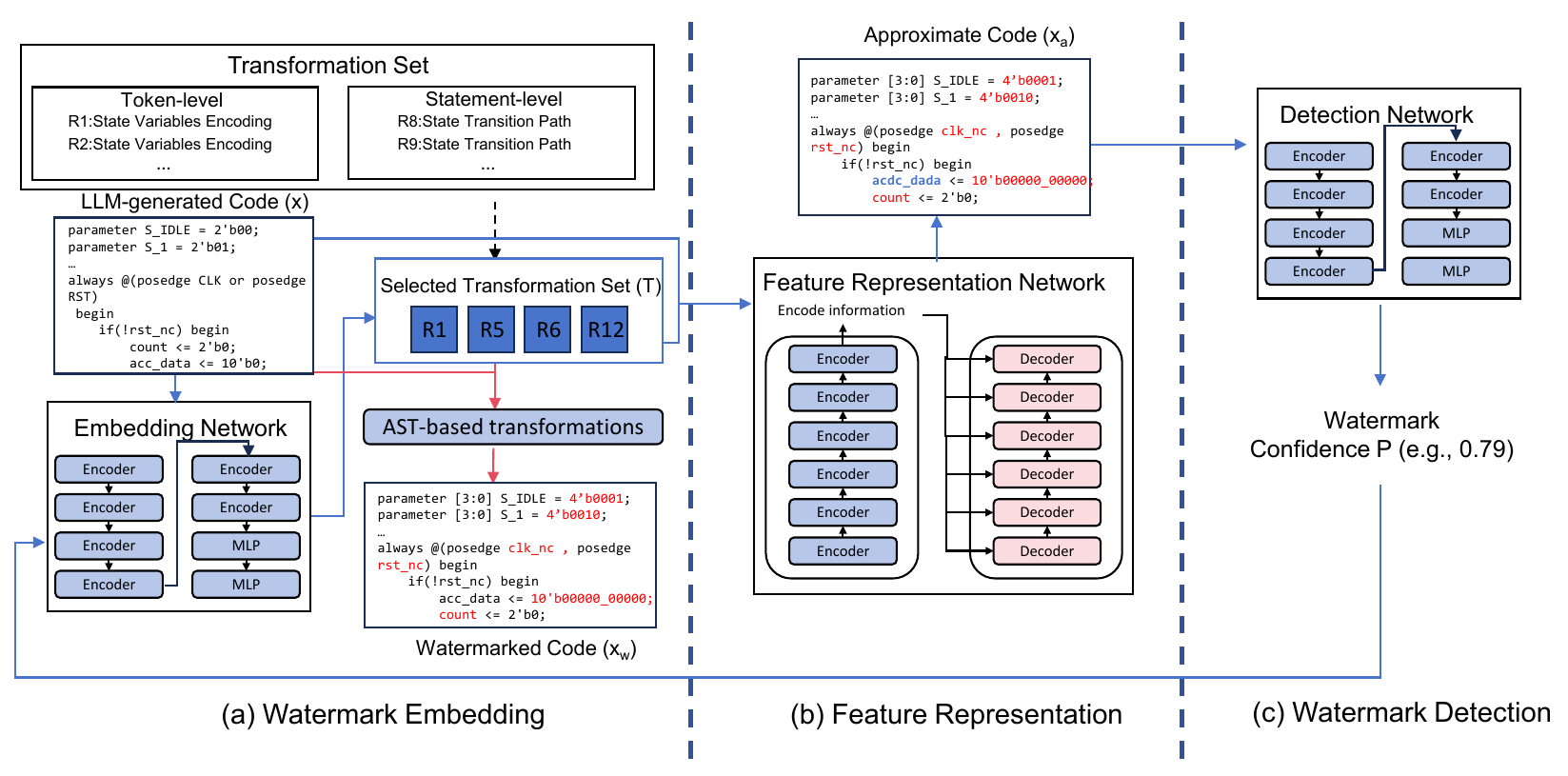}
    \caption{RTLMarker watermarking framework overview.}
    \label{fig:Overview}
\end{figure*}

%% file: formulation.tex
\section{PROBLEM FORMULATION}
In this section, we present a general formulation  of hardware watermarks. The embedding  of hardware watermarks can be viewed as a constrained optimization problem.  Specifically, we aim to find a suitable watermark $\delta$ to embed into the LLM-generated RTL code $x$, while satisfying constraints on effectiveness, robustness, and transparency, as follows:

\begin{equation}
\min_{\delta} \left( D(x, x + \delta) - {Det}_{rtl}(F(x + \delta)) - {Det}_{gate}(F(x + \delta)) \right)
\end{equation}

%\begin{equation}
%\min_{\delta}\! \left( D(x,x + \delta) \!-\! %\alpha \cdot  
%\text{Det}_{\text{rtl}}(F(x + \delta)) \!-\! %\beta \cdot 
%\text{Det}_{\text{gate}}(F(x + \delta)) \right)
%\end{equation}

 $D(x, x\,+\,\delta)$ represents the discrepancy between the LLM-generated  code $x$ and the watermarked code $x + \delta$. For transparency, this discrepancy should be minimized. $F(x+\delta)$ represents watermark attacks, e.g., variable name replacements. For robustness, the watermark should withstand such attacks. $Det_{rtl}$ and $Det_{gate}$ represent the 
watermark detection accuracy at the RTL and netlist level, respectively. % To comply with the effectiveness constraint,  the detection accuracy should be as high as possible. %$\alpha$ and $\beta$ are weighting coefficients used to balance the transparency and effectiveness of the watermark.

%The goal is to find suitable watermark $\delta$,
%This paragraph primarily defines the two main issues addressed in this paper through formal methods: the tradeoff between watermark robustness and concealment, and the dual watermarking mechanism.

%\warning{Every paragraph lacks : 1. Statement for the paragraph target in the first sentence. 2. Conclusion in the last sentence. 3. Every solution sentence should have a logic to let reviewers know why you do this, and what effect you expect though this solution.}

%\warning{Every sentence in the figure or table caption should be ended with .}
%\[
%\text{Minimize } \text{Distance}(x, x + \delta)
%\]
%\[
%\text{such that } \text{Detect\_RTL}(x + \delta) = \text{watermark} \quad \text{and} \quad \text{Detect\_Gate}(x + \delta) = \text{watermark}
%\]

%% file: method.tex
\section{METHOD}
\subsection{Overview}
RTLMarker's global flow is depicted in Figure \ref{fig:Overview}.
 In the training phase, RTLMarker comprises three principal modules: watermark embedding module,  feature representation module,
and watermark detection module. In the watermark embedding module, LLM-generated RTL code $x$ is fed into an embedding network, which subsequently produces  selected transformation set $T$ ( Line 6 in Algorithm \ref{alg:train}).
%Transformations within the set are ideally implemented via AST-based methods. Nevertheless, such AST-based methods are inherently non-differentiable, which impedes the opportunity for joint optimization of the embedding and detection networks.
Feature representation network  receives  the RTL code $x$ and the selected transformation set $T$ as inputs, employing a learning-based approach to approximate the code transformed by the rules from T (Line 9). The approximate code $x_a$ is then fed into a detection network to obtain the watermark confidence $P$ (Line 12) and  provide feedback to the embedding network. In the deployment phase,
after obtaining the  set T from the embedding network, we use the AST-based method with the rules in T to transform the code and obtain the watermarked code $x_w$, ensuring that the watermarked code is fully correct.
%code transformations are implemented by an AST-based method instead of utilizing the feature representation network.

%watermarks are embedded at two distinct granularities: RTL (Register Transfer Level) and Gate Level. Subsequently, the watermark detection module assesses the presence of watermarks in  llms-generated hardware at both the RTL and Gate levels.

\begin{algorithm}
\caption{RTLMarker Training Algorithm}
\label{alg:train}

\DontPrintSemicolon  % 避免每行末尾打印分号
 \textbf{Input:} $x$ LLM-generated code; $y_1$ appicable code transformations set for x; 
$y_2$ AST-transformed code ; $y_3$ label for watermark presence

 \textbf{Input:} $\theta_1$, $\theta_2$ and $\theta_3$: initial parameters for watermark embedding, feature representation, and watermark detection networks

 \textbf{Output:} \(\hat{\theta_1}\), \(\hat{\theta_2}\) and  \(\hat{\theta_3}\): trained parameters for watermark embedding, feature representation, and watermark detection networks

% \textbf{Initialize:} parameters $\theta_1$, $\theta_2$, and $\theta_3$ randomly\;
%\textbf{Define} learning rate $\eta$\;
 \textbf{Hyperparameters:} number of epochs $N$, 
 learning rate $\eta$\;

\For{epoch $=1$ \KwTo $N$}{
    
        \tcp{T: selected code transformations set.}

        $T \leftarrow \text{WatermarkEmbedding}(x; \theta_1)$\;
        $L_1 \leftarrow L_{embed}(T, y_1)$\;
        $g_1 \leftarrow \nabla_{\theta_1} L_1(\theta_1)$; 
        $\theta_1 \leftarrow \theta_1 - \eta \times g_1$\;
         \tcp{\(x_a\): approximate watermarked code.}
        $x_a\leftarrow \text{FeatureRepresention}((x , T); \theta_2)$\;
        $L_2 \leftarrow L_{fr}(x_a, y_2)$\;
        $g_2 \leftarrow \nabla_{\theta_2} L_2(\theta_2)$;
        $\theta_2 \leftarrow \theta_2 - \eta \times g_2$\;

        \tcp{P: confidence score.}
        $P \leftarrow \text{WatermarkDetection}(x_a; \theta_3)$\;
        $L_3 \leftarrow L_{detect}(P, y_3)$\;
        $g_3 \leftarrow \nabla_{\theta_3} L_3(\theta_3)$;
        $\theta_3 \leftarrow \theta_3 - \eta \times g_3$\;
    
}
\textbf{Return} \(\hat{\theta_1}\) = $\theta_1$, \(\hat{\theta_2}\) = $\theta_2$, \(\hat{\theta_3}\) =  $\theta_3$\;
\end{algorithm}

\subsection{Watermark Embedding}

\begin{table*}[htbp]
\centering

\caption{Verilog-specific code transformations.}
\label{tab:VerilogTrans}
\resizebox{\linewidth}{!}{
\begin{tabular}{ccccc} % 三列 'c'
\hline
Granularity & Description &  Transformation Rule \\ 
\hline
Token & State Variables Encoding & Using one-hot encoding instead of binary encoding, e.g., \enquote{RUN = 2'b10, STOP = 2'b11} to
\enquote{RUN = 3'b010, STOP = 3'b100}.  
   \\  
Token & Parameterized Module
 &  Bit widths can be parameterized. e.g., \texttt{a[0:32]} and \texttt{b[0:32]} can be parameterized as \texttt{a[0:\textit{width}]} and \texttt{b[0:\textit{width}]}.
 \\ 
Token & Base Conversion  &Binary, decimal, and hexadecimal can be converted to each other. \\ 
Token & Signal Sensitivity Formatting  & Signal sensitivity lists are reformatted. e.g., \enquote{(posedge clk1 or negedge clk2)} to   \enquote{(posedge clk1 , negedge clk2)}.\\
Token & Bit Separation  & Binary representations with many bits can include separators for better readability. \\
Token & Variable Name Replacement  & Replace variable names, e.g., the signal name $rst$ can be replaced with $rst\_nc$.\\
Token & Bit Order  &  Bit order can be reversed, e.g., \enquote{watermark[0:N]} to \enquote{watermark[N:0]}.  \\
Statement & State Transition Path  &Introduce specific transition sequences in the state transition path.\\
Statement & Combinational Logic Operation  &AND gates, OR gates, and NOT gates can be converted into each other.\\
Statement & Combinational Logic Assignment  &Add always@* during combinational logic assignment.\\
Statement & Ternary Operation  & If-else statements can be expressed using ternary operations.\\
Statement & Signal Initialization Order  &When initializing multiple signals, the order of initialization can be interchanged.\\
Statement & Add Comments  &Add unique comments to some simple signals. e.g., \enquote{//Input signal clk\_nc}.\\
Statement & Conditional Statement Order  &In the conditional statement \enquote{if(a \& b)}, the order of a and b can be interchanged.\\
Statement & Add Redundant Logic  &Add redundant logic that does not impact the normal execution of the code.\\
\hline
\end{tabular}
}
\end{table*}

\noindent\textbf{Rule-Based 
Verilog Code Transformations. }
%To address the gap in watermarking Verilog code generated by LLMs, 
we propose a set of Verilog-specific code transformations , based on equivalence transformation watermarking methodologies\cite{acm}, as shown in Table \ref{tab:VerilogTrans}.
The set includes 15 code transformation rules, categorized into two distinct granularities: token level and statement level. In curating this set,
we make efforts to avoid transformations\cite{acm} characterized by atypical styles that developers are unlikely to adopt, such as transforming $n<=2$ to $2>=n$.   
Furthermore, the applicability of these rules is constrained by specific contextual factors. For example, the state variable encoding rule can only be applied when Finite State Machine (FSM) is present in the code. 
Nevertheless, the process of watermark embedding does not necessitate the application of all transformation rules; rather, applying an adequate number to satisfy detection requirements suffices. Indeed, excessive utilization of transformation rules can adversely compromise the watermark's transparency.

%The set includes 15 code transformation rules which
%Verilog-specific code transformations are designed according to the syntax and semantic characteristics of 
% Verilog codes, offering strong concealment. In contrast, general code transformations, while applicable across various programming languages, exhibit poorer concealment.
%For example,
%the \fixme{transformation1}  semantically 
% converts the state variables of a finite state machine from one-hot to binary encoding, making it challenging for attackers to detect. In contrast, the \fixme{Transformation2}, which  converts "\texttt{n <= 2}" to "\texttt{2 >= n}", is more easily identifiable by attackers.
%However, Verilog-specific code transformations have a limited scope of application. The transformation1  is only applicable when FSMs are present in the code. In contrast, general code transformations have a broader application range. To ensure watermark robustness, We need to combine Verilog-specific code transformations with general code transformations.

\begin{figure}[htbp]
    \centering
    %\warning{In the figure, Code font should be \texttt{consolas}. Besides, the figure is vague. A clear code can be generated from pdf format PPT slides.}
    \includegraphics[width=0.9\linewidth]{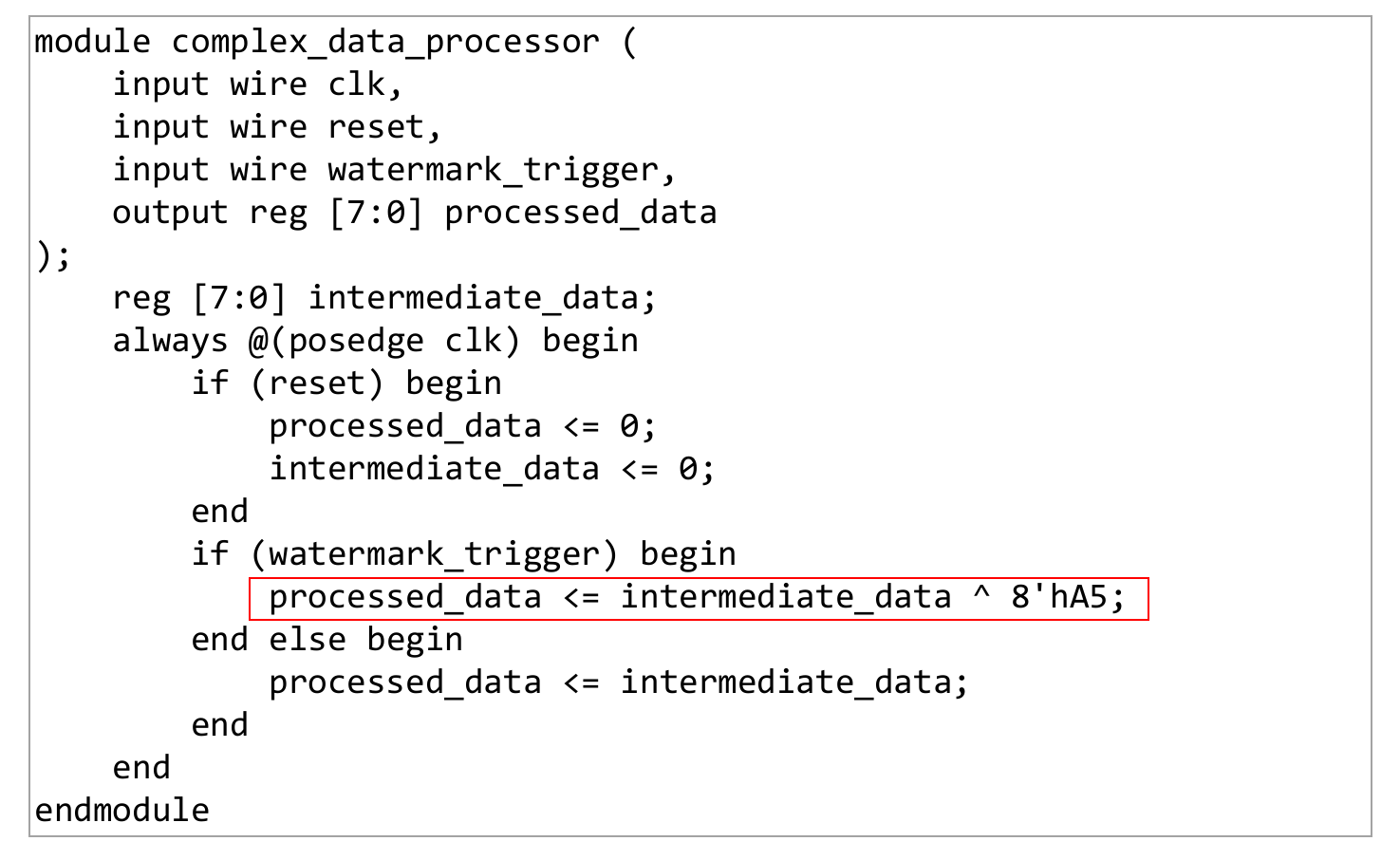}
    \caption{A watermark preserved in a synthesized netlist.}
    \label{fig:gate}
\end{figure}

Embedding watermarks at the Register Transfer Level (RTL) presents substantial challenges in achieving deep integration within the gate-level netlist, as most RTL  watermarks tend to dissipate  during the logic synthesis process.
We  discover that watermarks which modify the code's data flow and control flow exhibit greater resilience, often persisting in the gate-level netlist. However, modifications to the data flow often compromise the correctness of the code, thereby hindering the embedding of watermarks. In light of these observations, we consider integrating redundant control logic into the code's control flow to facilitate watermark embedding into the gate-level netlist. For example, as shown in Figure \ref{fig:gate},  we implement a redundant assignment logic that is activated exclusively when the $watermark\_trigger$ signal is asserted to 1. We encrypt the watermark information, which includes model and developer signatures, and use it as the right-hand side of the assignment statement. In this implementation, $8'hA5$ represents the encrypted watermark information. %c Note that this method is not suitable for all types of code. If the code has few lines or lacks control logic, applying this watermarking method  makes it easily removable.

 %We employ the Pyverilog \cite{pyverilog} for  code transformations. Initially, Verilog codes are parsed into an abstract syntax tree (AST). We then transform the code by adjusting nodes in the ASTs. Ultimately, the ASTs are converted back into Verilog code using Pyverilog’s \texttt{ASTCodeGenerator} class.

%\begin{comment}

%\end{comment}
\noindent\textbf{Learning-Based Watermark Embedding }
The goal of the watermark embedding network is to select suitable code transformations set from Table \ref{tab:VerilogTrans}, based on the LLM-generated code.
 %in accordance with the  LLMs-generated Verilog code $C$, while meeting the transparency requirements of the watermark. 
%Firstly, %the code is tokenized into discrete 
%tokens by a tokenizer. These tokens are then fed into transformer-based watermark embedding network, which outputs selected  transformation set.
The embedding network architecture incorporates multiple encoder layers to extract code features. The network's terminal segment consists of two fully connected layers, which are utilized for the regression prediction of the selected  transformation set $T$.
%Since the elements in the selected transformation set are discrete, being either 0 or 1, which makes them non-differentiable, we use Gumbel-Softmax\cite{AWT,jang2016categorical} to address this issue.
The network's loss function $L_{embed}$ comprises three parts: $L_{detect}$, $L_{mse}$ and $L_{trans}$, as described in Equ.\ref{equ:loss}, 
 where 
$L_{detect}$ 
  represents the confidence score output by the detection network, %identifying the presence of a watermark in the code. 
  $L_{mse}$   represents the mean squared error between the selected code transformations and the applicable code transformations, 
$L_{trans}$
  represents the number of selected code transformations. The coefficients $m$ and $n$ modulate the relative contributions of the various loss components in the optimization process.%This paper balances the strength and transparency of the watermark by constraining the number of code transformations .

\begin{equation}\label{equ:loss}
\begin{aligned}
    L_{embed} =  L_{detect} + m \cdot L_{trans} + n \cdot L_{mse}
\end{aligned}
\end{equation}

%\begin{equation}\label{equ:loss}
%\begin{aligned}
%    \text{$loss$} =  {L_{{conf}}} + m \cdot L_{\text{trans}} + n \cdot L_{\text{trans}}
%\end{aligned}
%\end{equation}

\subsection{Feature Represention}\label{sec:Feature Represention} Rule-based  code transformations are non-differentiable, posing a challenge for joint optimization between the watermark embedding and detection networks. To overcome this challenge, we develop a feature representation network which  
accepts  code $x$ and the selected  transformation set $T$  and generates feature approximate  watermarked code $x_a$. % , as shown in Equ.\ref{equ:seq2seq}.
%\begin{equation}\label{equ:seq2seq}
%\begin{aligned}
%watermarkedcode = FeatureRepresentation(tokens, set)
%\end{aligned}
%\end{equation}
The network comprises six encoder layers and  six decoder layers, following the transformer architecture\cite{vaswani2017attention}. 
The objective function employs cross-entropy loss, computed by comparing the network's output code tokens against those generated through Abstract Syntax Tree (AST)-based code transformations. While the network-generated code may occasionally exhibit character-level discrepancies, these anomalies seldom compromise the salient features of the embedded watermark.% In addition, we observe that the tokenizer exerts a subtle influence on the performance of model. For example, if the tokenizer does not recognize the keyword $endmodule$ in Verilog code, it mistakenly encodes it as separate tokens, $end$ and $module$. This incorrect segmentation disrupts the syntactic and semantic understanding of the code, resulting in a decline in model performance. 

\subsection{Watermark Detection}\label{sec:RTLDection} 
\textbf{Watermark Detection at the Register Transfer Level (RTL).} %Watermark detection at the  (RTL) primarily aims to determine the presence of a watermark within the code. 
The detection network ingests  the code generated by the  feature representation network . It  computes  a confidence score indicating the probability of a watermark's presence in the code. This score is then fed back to the watermark embedding network, creating a feedback loop. We include non-watermarked samples in training to enhance discrimination of unwatermarked code. The network optimizes the process using cross-entropy loss.
%We incorporate non-watermarked samples in the training dataset to improve the network's ability to identify unwatermarked code. The detection network utilizes cross-entropy loss for optimization.

\noindent\textbf{Watermark Detection at Netlist Level.} We  employ synthesis tools such as Yosys\cite{Yosys}  to perform logic synthesis on the LLMs-generated RTL code ,  generating the gate-level netlist. Subsequently, we trace the connections of the embedded watermark variable in the gate-level netlist. Through analysis of these connections, we identify the watermark's presence and characteristics. 
As shown in Figure \ref{fig:gate}, we  analyze  the connections  associated with $processed\_data$ variable in the gate-level netlist. 
Through this analysis,  we  extract the embedded watermark information, designated as 
 $8'hA5$, which allows us to further extract critical information, such as digital signatures of the model and developer.

%% file: evaluation.tex
\section{Evaluation}
\subsection{Experiment Setup}
\textbf{Target Model} We use the general purpose LLM, GPT-4\cite{gpt4} along with the hardware-specific  LLMs ChipGPT-FT\cite{chang2024data} and RTLCoder\cite{rtlcoder}, as our target models. We use RTLLM\cite{rtllm} and VerilogEval\cite{liu2023verilogeval}as benchmarks, which provide natural language descriptions for the generation of Verilog code.
%LLMs-generated Verilog codes are sourced from those provided in the RTLLM\cite{rtllm} and VerilogEval\cite{liu2023verilogeval} benchmark.

%\noindent\textbf{Datasets} we use Verilog code corpus provided in \cite{thakur2023benchmarking} as datasets.

\noindent\textbf{Baselines} We compare RTLMarker with WLLM\cite{wllm}, which is tailored for text watermarking, along with SWEET\cite{sweet} %and ACM\cite{acm},
which is specifically designed for general code watermarking. %As mentioned in Sweet, the entropy threshold for tokens is set to 1.2.

\noindent\textbf{Evaluation Metrics} We evaluate our watermarking framework from three aspects as follows.
\begin{itemize}[itemsep=2pt,topsep=0pt,parsep=0pt,partopsep=0pt,leftmargin=15pt]
    \item Effectiveness. The watermark should be successfully embedded and detected. To comprehensively evaluate our watermark framework, we employ Accuracy (ACC),  True Positive Rate (TPR) and False Positive Rate (FPR) as effectiveness metrics.
    \item Robustness. The watermark should be robust against typical attacks. We evaluate the robustness by measuring detection accuracy after variable name substitution attacks.%This paper also evaluates the robustness of the watermark based on the detection success rate after the watermarked code is synthesized into a logic gate netlist.
    \item Transparency. Transparency encompasses two metrics: the correctness rate of syntax and semantics in the watermarked code, and the number of code transformations utilized. Fewer transformations indicate better transparency.
    
\end{itemize}

\noindent\textbf{Implement Details}
 Our implementation of rule-based code transformations relies primarily on Pyverilog. Initially, verilog code is parsed into  an abstract syntax tree (AST). We then perform replacements of tokens and statements on the AST. The final step involves  generating the modified code from the updated AST. As Pyverilog lacks support for certain token-level rules, we supplement these transformations with methods based on regular expressions.   

\subsection{Effectiveness}
This experiment aims to assess RTLMarker's efficacy in discriminating between watermarked and non-watermarked code. The watermarked code is generated by RTLMarker, while the unwatermarked code is human-written and provided by the benchmark. Table \ref{tab:performance_comparison} illustrates the evaluation results for both RTLMarker and the baseline methods. Since WLLM and SWEET necessitate access to model-specific outputs,  particularly  logits, they are inherently unable to evaluate the closed-source GPT-4 model. To ensure the baselines achieve the best accuracy, we set the hardness parameter $\gamma$ to 3, the watermark tokens ratio $\delta$ to 0.25, the z-statistic score threshold to 4 and the entropy threshold to 0.6.

RTLMarker achieves an accuracy of over 95\% on the RTLLM benchmark and over 92\% on the VerilogEval benchmark at the Register Transfer Level (RTL), significantly outperforming the  baseline methods WLLM and SWEET. In the VerilogEval benchmark, the 156 Verilog problems sourced from the Hdlbits website encompass various complexities, including some  simple combinational circuits, which pose significant challenges for watermark embedding. Consequently, the True Positive Rate (TPR) observed on VerilogEval tends to be lower compared to that on RTLLM. 
Specifically, the TPR of RTLMarker on VerilogEval decreased by an average of $2.48\%$, in contrast to reductions of $28.78\%$ and $42.56\%$ for WLLM and SWEET, respectively.
%In addition, WLLM achieves a higher success rate than SWEET because SWEET excludes low-entropy tokens to improve the quality of the generated code, but this sacrifices the effectiveness of watermark. 
At the netlist level, RTLMarker achieves an accuracy of over 76\% on the RTLLM benchmark and over 59\% on the VerilogEval benchmark . Although the accuracy of  watermarking at the netlist level may not be exceptional, combining it with the RTL-level  watermarking can enhance both watermark embedding and detection performance.

%This effect is particularly pronounced in simple workloads like VerilogEval. 2.48  28.78 42.56

%However, the decline in accuracy on VerilogEval with our method is not particularly significant, as even simple workloads have applicable rules, albeit fewer in number. Additionally, our learning-based approach extracts features of the watermark, enhancing its detection capabilities.
\begin{table}[ht]
\centering
\caption{Evaluation for watermark effectiveness.}
\resizebox{\linewidth}{!}{
\label{tab:performance_comparison}
\begin{tabular}{@{}cccccc@{}}
\toprule
Method   & Benchmark & Model  & ACC(\%) & TPR(\%) & FPR(\%) \\ \midrule
\multirow{6}{*}{Ours(rtl)} & \multirow{3}{*}{RTLLM} & GPT-4     & 96.67 & 93.33 & 0 \\
                      &                        & RTLCoder & 95.00 & 90.00 & 0 \\
                      &                        & ChipGPT-FT  & 95.00 & 93.33 & 3.33 \\
                      & \multirow{3}{*}{VerilogEval}  & GPT-4     & 94.87 & 91.03 & 1.28 \\
                      &                        & RTLCoder & 92.62 & 88.46 & 3.2 \\
                      &                        & ChipGPT-FT  & 92.95 & 89.74 & 3.85 \\
\multirow{6}{*}{SWEET} & \multirow{3}{*}{RTLLM} & GPT-4     & --- & --- & --- \\
                      &                        & RTLCoder & 73.33 & 50 & 3.33 \\
                      &                        & ChipGPT-FT  & 71.67 & 46.67 & 3.33 \\
                      & \multirow{3}{*}{VerilogEval}  & GPT-4     & --- & --- & --- \\
                      &                        & RTLCoder & 58.65 & 20.51 & 3.21 \\
                      &                        & ChipGPT-FT  & 58.33 & 18.59 & 1.92 \\ 
\multirow{6}{*}{WLLM} & \multirow{3}{*}{RTLLM} & GPT-4     & --- & --- & --- \\
                      &                        & RTLCoder & 83.33 & 70 & 3.33 \\
                      &                        & ChipGPT-FT  & 86.67 & 76.67 & 3.33 \\
                      & \multirow{3}{*}{VerilogEval}  & GPT-4     & --- & --- & --- \\
                      &                        & RTLCoder & 63.14 & 29.49 & 3.21 \\
                      &                        & ChipGPT-FT  & 64.74 & 32.05 & 2.56 \\ 
\multirow{6}{*}{Ours(netlist)} & \multirow{3}{*}{RTLLM} & GPT-4     & 78.33 & 56.67 & 0 \\
                      &                        & RTLCoder & 76.67 & 53.33 & 0 \\
                      &                        & ChipGPT-FT  & 76.67 & 53.33 & 0 \\
                      & \multirow{3}{*}{VerilogEval}  & GPT-4     & 62.18 & 24.36 & 0 \\
                      &                        & RTLCoder & 62.82 & 25.64 & 0 \\
                      &                        & ChipGPT-FT  & 59.62 & 19.23 & 0 \\   \bottomrule                  
\end{tabular}
}
\end{table}
\subsection{Robustness}
In this experiment, we aim to evaluate whether  RTLMarker can effectively  withstand typical 
variable name replacement attack. We
consider renaming 25\%, 50\%, 75\% and 100\% of the
variables in the watermarked code, with new variable names ranging in length from 3 to 10 characters. Table \ref{tab:substitution} presents the evaluation results of RTLMarker. When the string replacement percentage is set to 100\%, the VerilogEval benchmark, due to its generally shorter lines of code, suffers  significant impacted , resulting in a decrease in  accuracy to 80.45\%. In contrast, the RTLLM is less affected, maintaining an  accuracy of 91.67\%. %Since variable name replacement is just one type of transformation in our rule-based code transformation set, it does not significantly impact our watermark's effectiveness.
Additionally, variable name replacement does not affect the accuracy of watermarking at the netlist level, but it does increase the complexity of watermark detection. Since we cannot rely on variable names which are replaced to analyze and extract watermark information, we need to analyze all variables with the same bit width and check for the presence of  watermark. %The probability of an 8-bit watermark variable coinciding with another 8-bit variable in the code is 0.003, so it almost does not affect the detection of the watermark.

\begin{table}[htbp]
\centering
\caption{Evaluation for variable name replacement attack.}
\resizebox{\linewidth}{!}{
\label{tab:substitution}
\begin{tabular}{@{}lcccccc@{}}
\toprule
\multirow{2}{*}{Attack} & \multicolumn{3}{c}{RTLLM} & \multicolumn{3}{c}{VerilogEval} \\ \cmidrule(lr){2-4} \cmidrule(lr){5-7}
           & ACC(\%) & TPR(\%) & FPR(\%) & ACC(\%) & TPR(\%) & FPR(\%) \\ \midrule
Attack@25\%     & 96.67  & 93.33 & 0 & 92.62  & 88.46 & 3.21 \\
Attack@50\%     & 96.67  & 93.33 & 0 & 90.38  & 86.53 & 5.76 \\
Attack@75\%     & 95.00  & 90.00 & 0 & 87.50  & 80.76 & 5.76 \\
Attack@100\%    & 91.67  & 83.33 & 0 & 80.45  & 70.51 & 9.62 \\ \bottomrule
\end{tabular}
}
\end{table}

%as well as baseline method. \fixme{DeepHardMark} achieves ACC results over \fixme{xx\%} in the RTLLM and VerilogEval benchmark. 

\subsection{Transparency}
\begin{figure}[htbp]
    \centering
    \includegraphics[width=0.8\linewidth]{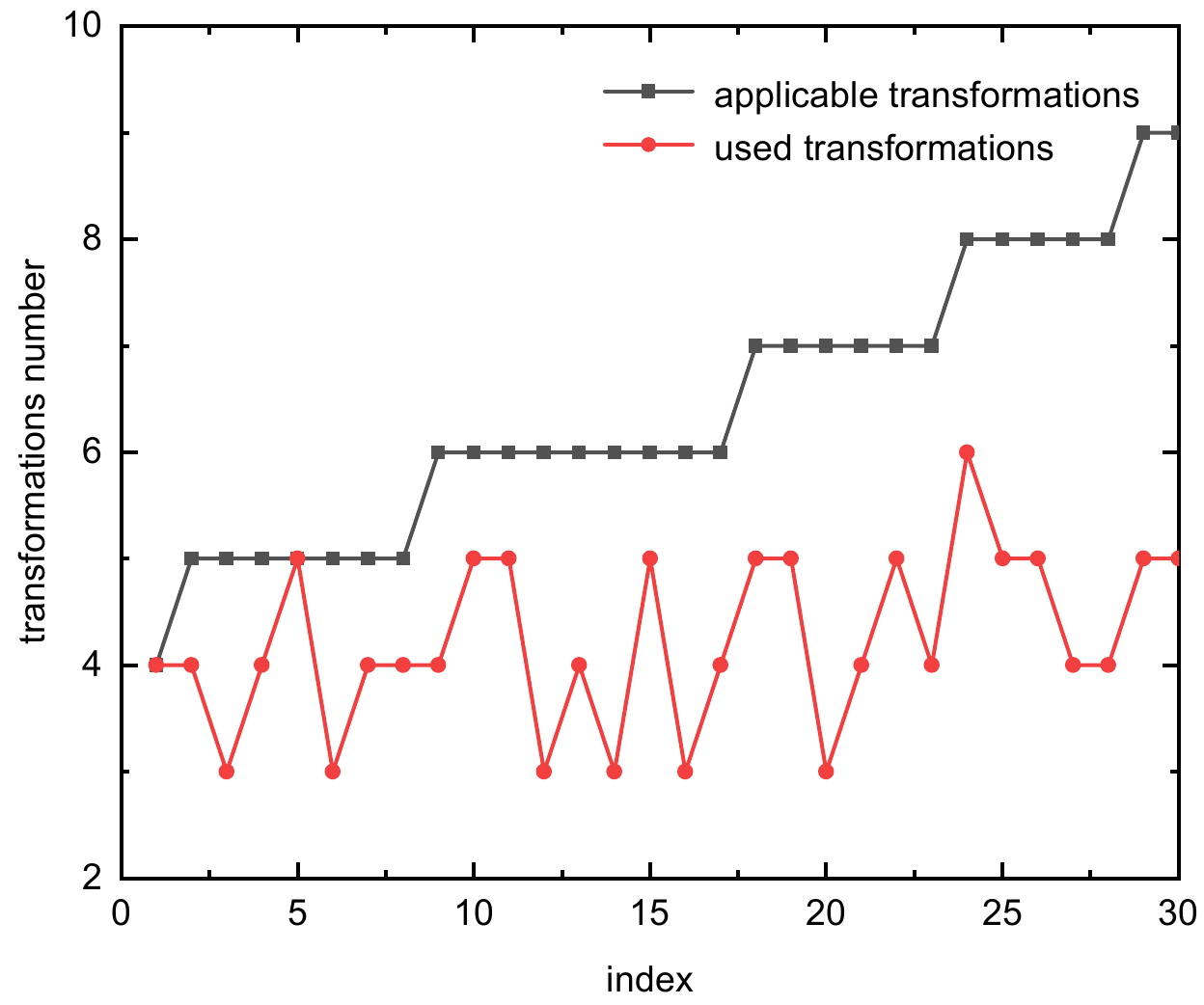}
    \caption{Comparison of the number of code transformations on the RTLLM benchmark.}
    \label{fig:trans}
\end{figure}

To evaluate the transparency of the watermark, we consider two aspects:  the impact of the watermark on the syntactic and semantic correctness of the RTL code and the number of code transformations utilized. Firstly, since the code transformation rules we designed are semantically equivalent, RTLMarker does not compromise  the correctness of syntax and semantics. Moreover, employing an excessive number of code transformations  implies substantial modifications to the code, potentially compromising the watermark's transparency. Consequently, it is crucial to apply the minimal number of code transformation  required to maintain effective watermark detectability. Figure \ref{fig:trans}  illustrates that the average number of applicable code transformations  in the RTLLM benchmark is $6.42$, while the number of code transformations  that RTLMarker utilizes is $4.25$,  effectively enhancing the transparency of the watermark. The x-axis (index) in the figure \ref{fig:trans}  represents different cases within the RTLLM benchmark.

%Ablation expiment \\

%% file: conclusion.tex
\section{CONCLUSION}
We present RTLMarker, a hardware watermarking framework to protect LLM-generated RTL Copyright. This paper proposes a rule-based Verilog code transformation set that ensures the functional correctness of the watermarked code while embedding the watermark into the RTL code and the synthesized netlist. This paper considers an inherent tradeoff between
watermark transparency and watermark effectiveness through the joint
optimization of the watermark embedding  and  detection network. RTLMarker demonstrates superior performance over existing watermarking framework in RTL code watermarking.

\begin{acks}
This paper is supported in part by National Key R\&D Program of China under grant 2023YFB4404400, and in part by the National Natural Science Foundation of China
(NSFC) under grant No.(62222411, 62025404, 62104229, 62104230).
\end{acks}